\documentclass[onecolumn,useAMS,articles,fleqn] {mnras}
\usepackage{epsfig}
\usepackage{amsmath}
\usepackage{graphicx}
\usepackage{amssymb}

\title[The extended law of star formation]{The extended law of star formation: the combined role of gas and stars}

\author[Dib et al.]{Sami Dib$^{1}$\thanks{E-mail: sami.dib@gmail.com}, Sacha Hony$^{2}$ Guillermo Blanc$^{3,4,5}$\\
$^{1}$Universidad de Atacama, Copayapu 485, Copiap\'{o}, Chile\\
$^{2}$Institut f\"{u}r Theoretische Astrophysik, Zentrum f\"{u}r Astronomie der Universit\"{a}t Heidelberg, Albert-\"{U}berle-Stra{\ss}e 2, 69120 Heidelberg, Germany.\\
$^{3}$Observatories of the Carnegie Institution for Science, 813 Santa Barbara St, Pasadena, CA, 91101, USA\\
$^{4}$Departamento de Astronom\'{i}a, Universidad de Chile, Camino del Observatorio 1515, Las Condes, Santiago, Chile\\
$^{5}$Centro de Astrof\'{i}sica y Tecnolog\'{i}as Afines (CATA), Camino del Observatorio 1515, Las Condes, Santiago, Chile\\
}

\begin{document}
\maketitle

\date{Accepted XXX. Received XXX}

\pagerange{\pageref{firstpage}--\pageref{lastpage}}
\pubyear{2016}
\label{firstpage}

\begin{abstract} 

We present a model for the origin of the extended law of star formation in which the surface density of star formation ($\Sigma_{\rm SFR}$) depends not only on the local surface density of the gas ($\Sigma_{g}$), but also on the stellar surface density ($\Sigma_{*}$), the velocity dispersion of the stars, and on the scaling laws of turbulence in the gas. We compare our model with the spiral, face-on galaxy NGC 628 and show that the dependence of the star formation rate on the entire set of physical quantities for both gas and stars can help explain both the observed general trends in the $\Sigma_{g}-\Sigma_{\rm SFR}$ and $\Sigma_{*}-\Sigma_{\rm SFR}$ relations, but also, and equally important, the scatter in these relations at any value of $\Sigma_{g}$ and $\Sigma_{*}$. Our results point out to the crucial role played by existing stars along with the gaseous component in setting the conditions for large scale gravitational instabilities and star formation in galactic disks. 

\end{abstract} 

\begin{keywords}
galaxies: star formation - galaxies:  kinematics and dynamics - galaxies: stellar content - ISM: structure - galaxies: ISM - galaxies: evolution
\end{keywords}

\section{INTRODUCTION}\label{motiv}

The star formation rate (SFR) is the quantity that describes how galaxies convert their gas reservoirs into stars per unit time. Quantifying the dependence of the SFR on the global properties of galaxies as well as on the local conditions within galaxies is essential towards understanding their observed properties and their dynamical and chemical evolution across cosmic time. Traditionally, observational studies have sought the correlation between the surface density of star formation ($\Sigma_{\rm SFR}$) and the surface density of the gas $\Sigma_{g}=\Sigma_{\ion{H}{i}}+\Sigma_{{\rm H_{2}}}$, where $\Sigma_{\ion{H}{i}}$ and $\Sigma_{{\rm H_{2}}}$ are the surface densities of the neutral and molecular hydrogen, respectively. The emerging picture from all of these works is that $\Sigma_{\rm SFR} \propto \Sigma_{g}^{n}$ with $n \approx 1.4$  (e.g., Schmidt 1959; Kennicutt 1998; Bigiel et al. 2008; Blanc et al. 2009). Other studies found that the surface density of star formation scales linearly or sub-linearly ($n \lesssim 1$) with the surface density of molecular hydrogen traced by CO lines or with the surface density of molecules that trace higher density gas such as HCN (e.g., Gao \& Solomon 2004; Shetty et al. 2013; Liu et al. 2016). Several ideas have been proposed in order to explain the origin of the star formation scaling relations. The earliest scenarios proposed that stars form as a result of gravitational instabilities (GI) in the gaseous component of galactic disks over a timescale which is the local free-fall time of the gas and which is given by $t_{ff,g} \propto \rho_{g}^{-0.5}$, where $\rho_{g}$ is the local gas volume density. For a constant scale height of the disk, $\rho_{g} \propto \Sigma_{g}$ and thus $\Sigma_{\rm SFR} \propto \Sigma_{g}/t_{ff,g} \propto \Sigma_{g}^{1.5}$ (e.g., Madore 1977). Wong \& Blitz (2002) argued that the value of the star formation law slope is related to the value of the molecular fraction $f_{{\rm H_{2}}}=\Sigma_{{\rm H_{2}}}/\Sigma_{g}$ and Blitz \& Rosolowsky (2006) showed that $f_{{\rm H_{2}}}$ can be related to the pressure of the interstellar medium. It was also suggested that the value of $n$ is related to the width of the density probability distribution function of the interstellar gas and to the threshold density that is associated with the gas tracer (Tassis 2007; Wada \& Norman 2007). Escala (2011) argued that a correlation exists between the largest mass-scale for structures not stabilised by rotation and the SFR. Other groups (e.g., Krumholz \& McKee 2005; Padoan \& Nordlund 2011; Hennebelle \& Chabrier 2011; Federrath 2013; Kraljic et al. 2014) explored ideas based on the role of turbulent fragmentation in GMCs and in which the SFR is a function of the dynamical properties of the clouds. Meidt et al. (2013) argued that the star formation rate in molecular clouds in M51 may correlate with the intensity of the dynamical pressure the clouds are subjected to. The role of feedback coupled to turbulent fragmentation and its effects on the regulation of the SFR on galactic scales have been included in a number of models (e.g., Dopita 1985; Dopita \& Ryder 1994; Dib et al. 2011a,b; Dib 2011a,b; Renaud et al. 2012; Dib et al. 2013; Orr et al. 2017). 

It is however necessary to include stars in the treatment of GI on large scales in galactic disks, since in most disk galaxies, the stellar surface density is observed to be a factor $\approx 10-100$ larger than the gas surface density (e.g., Leroy et al. 2008). The role of existing stars in determining the development of gravitational instabilities has been investigated in a limited number of studies. Jog \& Solomon (1984a,b) explored the characteristics of the gravitational instability in a two fluid medium (gas and stars) in which both components interact gravitationally with each other and are treated each as an isothermal gas with specific velocity dispersions. One of their main conclusions is that even when each fluid component is gravitationally stable, the joint fluid system may be gravitationally unstable. Rafikov (2001) expanded the study of Jog \& Solomon to the case where the stars are treated as a collisionless  component. Setting stars aside, Romeo et al. (2010) investigated the role of turbulent motions on the stability of galactic disks. They described interstellar turbulence using scaling laws that relate the size of a region to the gas surface density ($\Sigma_{g}$) and gas velocity dispersion ($\sigma_{g}$). Romeo \& Agertz (2014) investigated the development of GI for various regimes of turbulence (i.e., different dependence of $\Sigma_{g}$ and $\sigma_{g}$ on the physical scale). In parallel, Romeo \& Wiegert (2011) and Romeo \& Falstad (2013) proposed a derivation of the effective Toomre $Q$ parameter (Toomre 1964) for multicomponent disks of stars and gas and taking into account the effects of disk thickness. Shadmehri \& Khajenabi (2012) and Hoffman \& Romeo (2012) coupled aspects of the analysis of Jog \& Solomon (1984a) to that of Romeo et al. (2010) and investigated the linear growth rate of the GI in a gas+star galactic disk while at the same time accounting for the turbulent nature of the gas. On the observational side, Shi et al. (2011) showed that the scatter in the $\Sigma_{g}-\Sigma_{\rm SFR}$ relation may be reduced if $\Sigma_{\rm SFR}$ is a function that depends on both $\Sigma_{g}$ and $\Sigma_{*}$. When describing $\Sigma_{\rm SFR}$ as the product of two power law functions of the gas and stellar surface densities ($\Sigma_{\rm SFR} \propto \Sigma_{g}^{\alpha}~\Sigma_{*}^{\beta}$). They obtained $\alpha=0.8\pm0.01$ and $\beta=0.63\pm0.01$ from the combined measurements on sub-galactic scales (scales of $\approx 750$ pc) of 12 nearby galaxies, with a non-negligible galaxy-to-galaxy scatter when the data of each galaxy is fitted individually (see aslo Westfall et al. 2014). Rahmani et al. (2016) performed a similar study for the Andromeda galaxy, and showed that these exponents may well depend on the distance from the centre of the galaxy. It is important to mention that the description of the extended law of star formation as being the product of two power-laws (for gas and stars) is an empirical one, and possibly is an over-simplification of the physical processes that may be connecting the gas and stellar properties to the star formation rate. 

However, in all of these above mentioned works, the origin of the dependence of the surface density of star formation on the local properties of the gas and stars has not been explicitly quantified. In this work, we examine the role of GI in a two fluid medium (gas and stars) and investigate the quantitative relationship between the surface density of the star formation rate and the surface densities and velocity dispersions of the stellar and turbulent gaseous components\footnote{Keeping with the terminology used in Shi et al. 2011, we also use the term "extended" to describe the dependence of the star formation rate on physical quantities pertaining to both gas and stars in galactic disks}. The basis of our model is that the fastest growing mode of the GI is the one that is directly connected to the star formation rate. In \S~\ref{framework} we recall the basic equations that lead to the derivation of the wavelength of the fastest growing mode in a stellar+turbulent gas disk ($\lambda_{\rm SF}$), and to the quantitative dependence of $\Sigma_{\rm SFR}$ on $\lambda_{\rm SF}$ and other gas and stellar structural and dynamical properties. In \S~\ref{applicationngc628} we make a detailed comparison between the predictions of the $\Sigma_{\rm SFR}$ from our model and the observed values for the face-own, spiral galaxy NGC 628. We also discuss how including the effects of stellar feedback can affect, and in fact improve, the matching between the models and the observations. In \S~\ref{conclusions} we conclude.

\section{Theoretical framework}\label{framework}

\subsection{Derivation of the most unstable mode}\label{mostunstable}

The initial analytical formalism follows that of Jog \& Solomon (1984a) for the two fluid approach, Romeo et al. (2010) concerning the inclusion of the turbulent motions of the gas, and Shadmehri \& Khajenabi (2012) who combined both aspects. We recall here some of the basic assumptions. Both gas and stars in the disk are treated as isothermal fluids with velocity dispersions $\sigma_{g}$ and $\sigma_{*}$ and their unperturbed surface densities are given by $\Sigma_{g}$ and $\Sigma_{*}$, respectively. The scale-heights of the gaseous and stellar components are given by $h_{g}$ and $h_{*}$, respectively. Starting from the perturbed and coupled hydrodynamical gas-stars equations, and a solution for the perturbed quantities that has the functional form $\rm {exp}\left[i \left(k r+\omega t \right)\right]$, Jog \& Solomon (1984a) derived the dispersion relation that describes the growth rate of the instability in the linear regime, $\omega$. This is given by the following biquadratic equation:  

\begin{equation}
{\omega}^{4}-{\omega}^{2}\left(\alpha_{*}+\alpha_{g} \right)+ \left(\alpha_{*}\alpha_{g}-\beta_{*}\beta_{g} \right)=0,
\label{eq1}
\end{equation}

\noindent where 

\begin{equation}
\alpha_{*} = \kappa^{2}+k^{2} \sigma_{*}^{2} - 2\pi G k \Sigma_{*} \frac{1}{1+k h_{*}}, 
\label{eq2}
\end{equation}

\begin{equation}
\alpha_{g} = \kappa^{2}+k^{2} \sigma_{g}^{2} - 2\pi G k \Sigma_{g} \frac{1}{1+k h_{g}},
\label{eq3}
\end{equation}

\begin{equation}
\beta_{*} = 2\pi G k \Sigma_{*} \frac{1}{1+k h_{*}}
\label{eq4}
\end{equation}

\begin{equation}
\beta_{g} = 2\pi G k \Sigma_{g}  \frac{1}{1+ k h_{g}},
\label{eq5}
\end{equation}

\noindent where $\kappa$ is the epicyclic frequency, and $1/(1+k h_{g})$ and  $1 / (1+ k k_{*})$ are the reduction factors due to the gas and stellar disks thickness, respectively (Vandevoort 1970; Romeo 1992). The solutions to Eq.~\ref{eq1} are given by

\begin{equation}
\omega^{2}\left(k\right)=\frac{1}{2}\left[\left(\alpha_{*}+\alpha_{g}\right)  \pm \sqrt{\left(\alpha_{*}+\alpha_{g}\right)^{2}-4\left(\alpha_{*}\alpha_{g}-\beta_{*}\beta_{g}\right)} \right]. 
\label{eq6}
\end{equation} 

Only one of these roots allows for unstable modes to grow. This is given by

\begin{equation}
\omega_{-}^{2}\left(k\right)=\frac{1}{2}\left[\left(\alpha_{*}+\alpha_{g}\right) - \sqrt{\left(\alpha_{*}+\alpha_{g}\right)^{2}-4\left(\alpha_{*}\alpha_{g}-\beta_{*}\beta_{g}\right)} \right]. 
\label{eq7}
\end{equation}

inserting back the expressions of $\alpha_{*}$, $\alpha_{g}$, $\beta_{*}$, $\beta_{g}$ from Eqs.~\ref{eq2}-\ref{eq5} into Eq.~\ref{eq7}  and working in the limit where $h_{*} k \lesssim 1$ and $h_{g} k \lesssim 1$ i.e., in the limit of the thin disk approximation in which case the perturbations have a length-scale that is of the order, or larger, than the gaseous and stellar scales heights, then Eq.~\ref{eq7} becomes:

\begin{align}
 \omega_{-}^{2}\left(k\right) &= \kappa^{2} + \frac{\left(\sigma_{*}^{2}+\sigma_{g}^{2}\right)}{2} k^{2} - \pi G \left(\Sigma_{*}+\Sigma_{g}\right) k  \nonumber \\ 
& - \frac{k}{2} {\biggr[} \left(\sigma_{*}^{2}-\sigma_{g}^{2}\right)^{2} k^{2}+ 4 \pi G \left(\sigma_{*}^{2}-\sigma_{g}^{2}\right)\left(\Sigma_{g}-\Sigma_{*}\right) k+ 4 \pi^{2} G^{2} \left(\Sigma_{*}+\Sigma_{g}\right)^{2} {\biggr]^{1/2}},
\label{eq8}
\end{align}  

\noindent which is independent of both $h_{*}$ and $h_{g}$. While this assumption is not explicitly necessary if $h_{g}$ and $h_{*}$ are known, the advantage of applying the thin disk approximation is that these two scales heights are generally poorly constrained for face-on disk galaxies. Under the plausible assumption that the gaseous component is turbulent, the surface density and velocity dispersion of the gas are scale dependent and are assumed to follow Larson type scaling relations (Larson 1981), and are given by:

\begin{equation}
\Sigma_{g} = \Sigma_{g0} \left(\frac{k}{k_{0}}\right)^{-a}, 
\label{eq9}
\end{equation}

and 

\begin{equation}
\sigma_{g} = \sigma_{g0}\left(\frac{k}{k_{0}}\right)^{-b},
\label{eq10}
\end{equation}

\noindent where $a$ and $b$ are descriptive of the nature of turbulent motions, and $\Sigma_{g0}$ and $v_{g0}$ are the surface density and velocity dispersion on the scale of the spatial resolution of the observations (i.e., $2\pi/k_{0}$), respectively. Replacing Eqs.~\ref{eq9}-\ref{eq10} in Eq.~\ref{eq8} yields

\begin{align}
\omega_{-}^{2}\left(k\right) &= \kappa^{2} + \frac{\sigma_{*}^{2}}  {2} k^{2}+ \frac {\sigma_{g0}^{2}} {2} \left(\frac{k}{k_{0}}\right)^{-2b} k^{2} 
- \pi G \left(\Sigma_{*}+\Sigma_{g0}\left(\frac{k}{k_{0}}\right)^{-a}\right) k  \nonumber \\ 
& - \frac{k}{2} {\biggr[} \left(\sigma_{*}^{2}-\sigma_{g0}^{2} \left(\frac{k}{k_{0}}\right)^{-2 b}\right)^{2} k^{2}+ 4 \pi G \left(\sigma_{*}^{2}-\sigma_{g0}^{2} \left(\frac{k}{k_{0}}\right)^{-2b}\right)\left(\Sigma_{g0} \left(\frac{k}{k_{0}}\right)^{-a} - \Sigma_{*}\right) k+ \nonumber \\ 
& 4 \pi^{2} G^{2} \left(\Sigma_{*}+\Sigma_{g0} \left(\frac{k}{k_{0}}\right)^{-a}\right)^{2} {\biggr]^{1/2}}.
\label{eq11}
\end{align}  

We posit that the fastest growing mode is directly linked to the star formation rate. The fastest growing mode, $k_{\rm SF}$, can be obtained by requiring that 

\begin{equation}
\frac{d \omega_{-}^{2}  \left(k_{\rm SF}\right)} {d k}=0,
\label{eq12}
\end{equation}

which is an equation that can be solved numerically. It is interesting to note that Eq.~\ref{eq12} possesses always a positive, non-zero root, for any positive values of the exponents $a$ and $b$ when $ a < 1/2$ and $b < 1/2$. These values are the typical upper limits measured for  $a$ and $b$ in all phases of the interstellar gas. The full analytical expression of Eq.~\ref{eq12} is of little direct interest here and is given in Appendix~\ref{appa}.  It also implies that the SFR is independent of the galactic rotation (i.e., no dependence of $\kappa$). This is consistent with the findings of Dib et al. (2012) who found no correlation between the star formation levels in Galactic molecular clouds and the degree of shear the clouds are subjected to. Following the method of Dib et al. (2012), Thilliez et al. (2014) reached a similar conclusion for molecular clouds in the Large Magellanic Cloud. It is useful to point out that our definition of the characteristic length scale of the most unstable mode ($\lambda_{\rm SF}=2\pi/k_{\rm SF}$), which we associate with star formation, is different from the one used by Romeo \& Falstad (2013) (see also Fathi al. 2015). The latter authors define the characteristic length scale as being the scale at which the effective Toomre parameter drops below unity.  

\subsection{Connection to the SFR}\label{connectionsfr}

The SFR can be directly related to the length scale of the most unstable mode $\lambda_{\rm SF}$. The mass of the gas that is associated with $\lambda_{\rm SF}$ is given by:

\begin{equation}
M_{\rm SF} = \bar{\rho} V_{\rm SF}, 
\label{eq13}
\end{equation}

where $\bar{\rho}$ is the average density within the mass $M_{\rm SF}$, and $V_{\rm SF}$ is the volume of the gravitationally unstable gas. In the limit of $k_{\rm SF} h_{g}  \lesssim 1$  as adopted above, $V_{\rm SF}$ is given by $V_{\rm SF}=\pi \lambda_{\rm SF}^{2} 2h_{g}$ and the volume density of the gas can be replaced by the gas surface density. Thus, Eq.~\ref{eq13} becomes: 

\begin{equation}
M_{\rm SF} = \frac{\Sigma_{g}} {2 h_{g}} \pi \lambda_{\rm SF}^{2} 2h_{g} = \Sigma_{g} \pi \lambda_{\rm SF}^{2}.
\label{eq14}
\end{equation} 

The theoretical star formation rate is then given by:

\begin{equation}
{\rm SFR}_{th} \approx \epsilon_{ff} \frac{M_{\rm SF}} {t_{ff}}, 
\label{eq15}
\end{equation}

\noindent where $t_{ff}$ is the free fall time of the unstable mass reservoir, and $\epsilon_{ff}$ is the efficiency of the star formation process per unit free fall time. We approximate $t_{ff}$ with $1/\sqrt{G \rho_{mp0}}$, where $\rho_{mp0}$ is the gas volume density at the mid-plane. The mid-plane volume density can be written as (e.g., Krumholz \& McKee 2005): 

\begin{equation}
\rho_{mp0} \approx \frac{\pi G \phi_{P} \Sigma_{g0}^{2}}{2 \sigma_{g0}^{2}}, 
\label{eq16}
\end{equation}

\noindent where $\Sigma_{g0}$ and $\sigma_{g0}$ carry the same meaning as in \S.~\ref{mostunstable} and with $\phi_{P}$ being a term of order unity that describes the contribution of stars to the mid plane pressure. An approximation of $\phi_{P}$ is given by (e.g., Elmegreen 1989):

\begin{equation}
\phi_{P} \approx 1+\frac{\Sigma_{*}}{\Sigma_{g0}} \frac{\sigma_{g0}}{\sigma_{*}}.
\label{eq17}
\end{equation}

With these approximations $t_{ff}$ can be written as: 

\begin{equation}
t_{ff} = \sqrt{\frac{2}{\pi}} \frac{1}{G} \frac{\sigma_{g0}}{\Sigma_{g0}} \left(1+\frac{\Sigma_{*}}{\Sigma_{g0}}\frac{\sigma_{g0}}{\sigma_{*}}\right)^{-1/2}.
\label{eq18}
\end{equation} 

Combining Eq.~\ref{eq14} and Eq.~\ref{eq18} yields the expression for the ${\rm SFR}_{th}$:

\begin{equation}
{\rm SFR}_{th} = \epsilon_{ff} \frac{\pi^{3/2}} {2^{1/2}} G \lambda_{\rm SF}^{2} \frac{\Sigma_{g0}^{2}}{\sigma_{g0}} \left(1+\frac{\Sigma_{*}}{\Sigma_{g0}}\frac{\sigma_{g0}}{\sigma_{*}}\right)^{1/2}.
\label{eq19}
\end{equation}

\begin{figure}
\begin{center}
\epsfig{figure=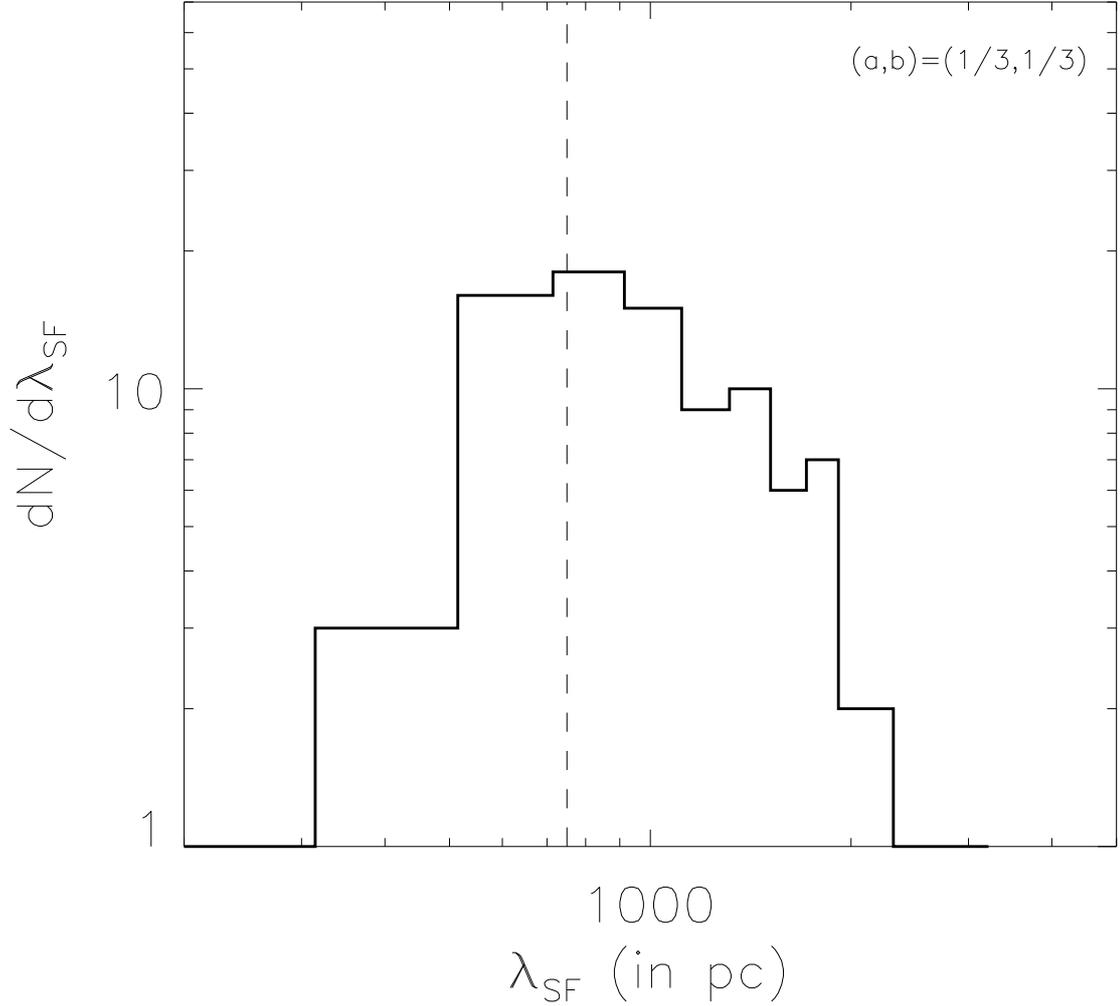,width=\columnwidth}
\end{center}
\caption{Distribution function of the wavelength of the most unstable mode $\lambda_{\rm SF}$ for the sample of data points that are used in this study (see text for the selection criteria) and for values of $a=1/3$ and $b=1/3$. The spatial resolution of the observations ($\lambda_{0}=750$ pc) is shown with the dashed line.}
\label{fig1}
\end{figure}

\begin{figure}
\begin{center}
\epsfig{figure=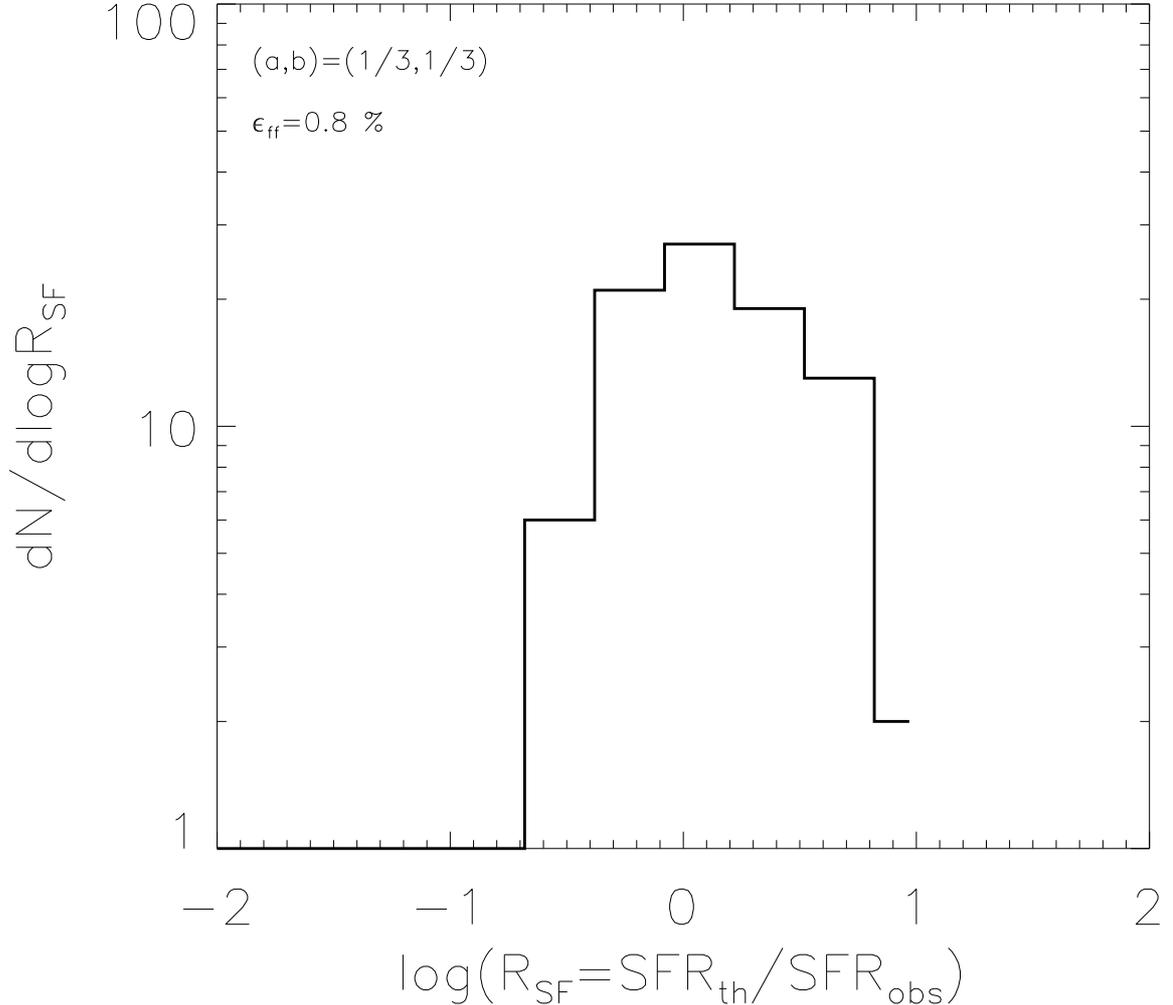,width=\columnwidth}
\end{center}
\caption{Distribution function of the ratio of the theoretical estimate of the star formation rate (${\rm SFR}_{th}$) to the observed one (${\rm SFR}_{obs}$) for the sample of data points that are used in this study (see text for the selection criteria) and for values of $a=1/3$, $b=1/3$, and $\epsilon_{ff}=0.8 \%$.}
\label{fig2}
\end{figure}

The theoretical estimate of the surface density of the star formation rate is then simply given by:

\begin{equation}
\Sigma_{{\rm SFR},th}=\frac{{\rm SFR}_{th}} {S},  
\label{eq20}
\end{equation}

\noindent where $S$ is the surface area covered by the beam size in the observations. 

\section{Application to NGC 628}\label{applicationngc628}

\begin{figure}
\begin{center}
\epsfig{figure=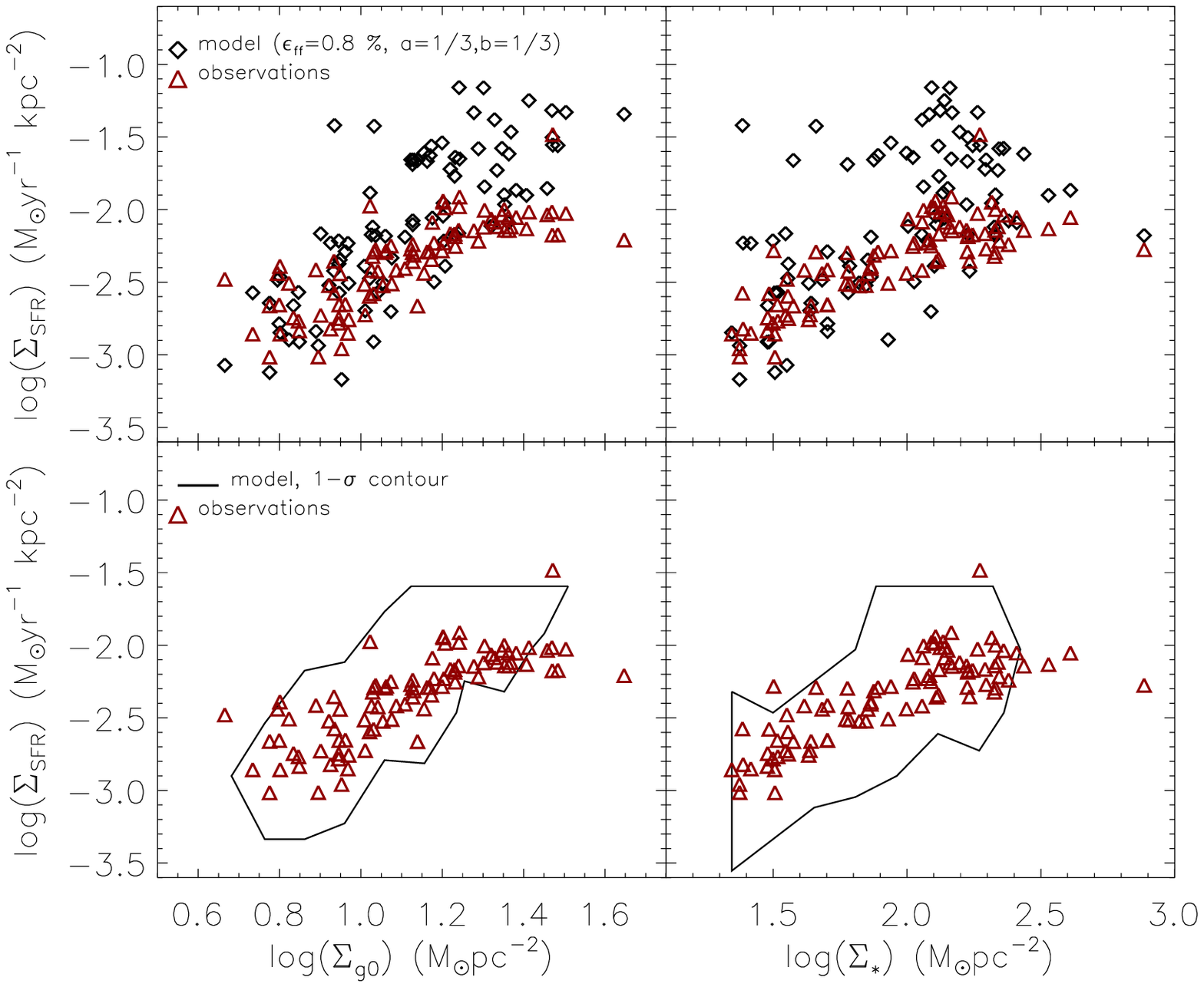,width=0.80\columnwidth}
\end{center}
\vspace{1cm}
\caption{The surface density of the star formation rate plotted as a function of the total gas surface density (left panels) and stellar surface density (right panels). The observational data are shown with the red open triangles. The theoretical estimates from the model are shown with the open black diamonds (top panels), and as a closed contour containing $68\%$ of the estimates (bottom panels). The free parameters of the model are taken to be $a=b=1/3$ and $\epsilon_{ff}=0.8\%$. Here $\Sigma_{g0}$ is the total surface density of the gas measured on a spatial scale which is equal to the spatial resolution of the observations (i.e., 750 pc).}
\label{fig3}
\end{figure}

 \begin{figure}
\begin{center}
\epsfig{figure=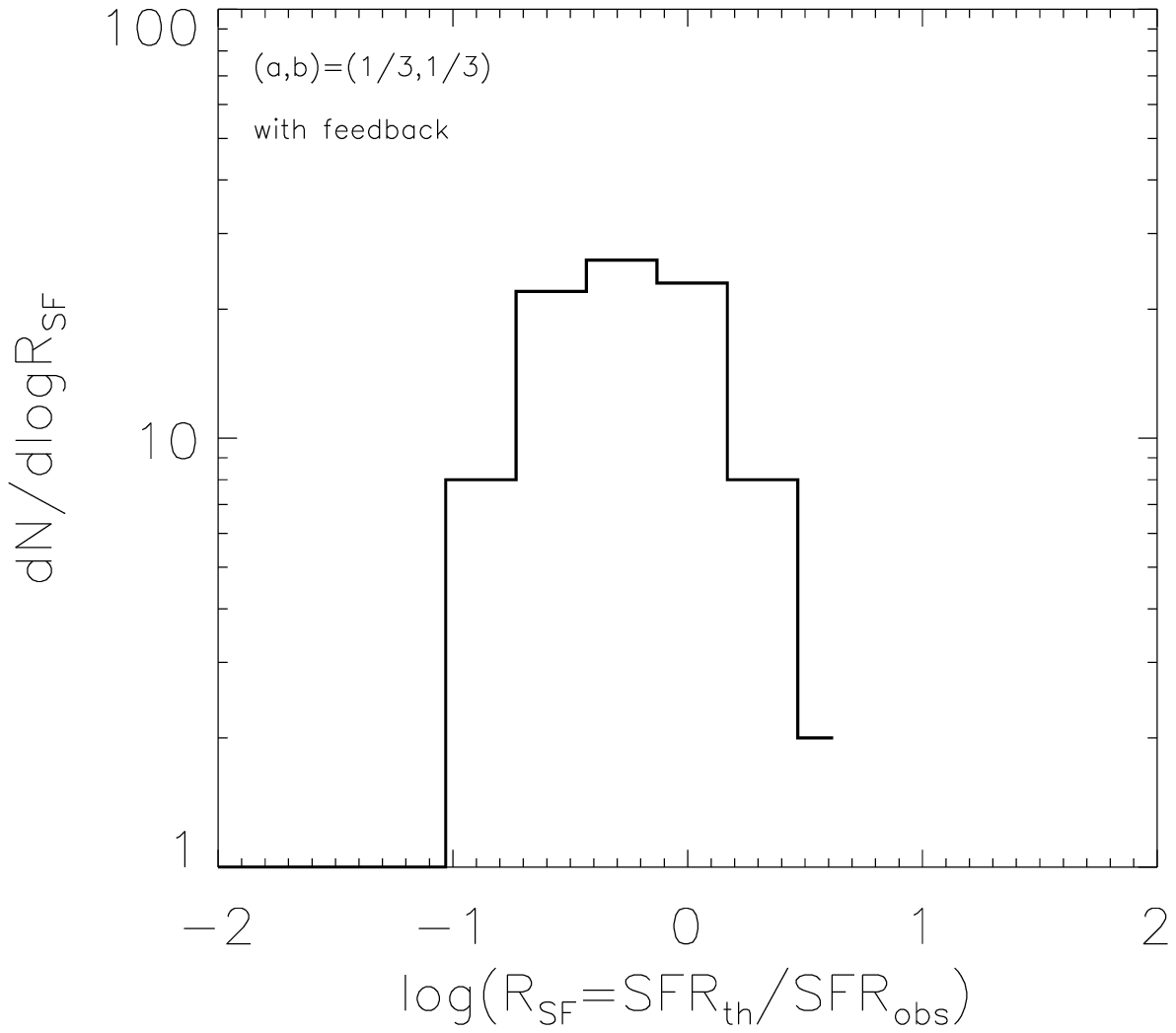,width=\columnwidth}
\end{center}
\caption{Similar to Fig.~\ref{fig2}, but in this case the efficiency of star formation per unit free-fall time is taken to depend on the gas surface density following $\epsilon_{ff}\propto \Sigma_{g}^{-0.34}$ and normalised to be $0.8\%$ at $\Sigma_{g}=1$ M$_{\odot}$ pc$^{-2}$.}
\label{fig4}
\end{figure}

We test our model by comparing its predictions to the face-on, spiral galaxy NGC 628.  The values of $\Sigma_{\ion{H}{i}}$ and $\sigma_{\ion{H}{i}}$ for NGC 628 are derived from the moment 0 and moment 2 maps of the \ion{H}{i} Nearby Galaxy Survey (THINGS; Walter et al. 2008). The spatial resolution (i.e., beam size) for these observations at the distance of NGC 628 are 750 pc, thus, the surface area of the resolution element in NGC 628 used in this work is $S=750\times750$ pc$^{2}$. As in Shi et al. (2011), the values of $\Sigma_{{\rm H_{2}}}$ are derived from the moment 0 CO $J=1-0$ BIMA SONG survey (Helfer et al. 2003), and the stellar surface density is taken from the SIRTF Nearby Galaxies Survey (SINGS; Kennicutt et al. 2003). Since the \ion{H}{i} gas is ubiquitously present in the galaxy, we approximate the velocity dispersion of the gas as being the velocity dispersion of the \ion{H}{i} gas, $\sigma_{g} \approx \sigma_{\ion{H}{i}}$. Measurements of the stellar velocity dispersions in nearby galaxies are scarce. Yet, the VENGA survey has made such measurements, with selected mosaics, for a sample of nearby galaxies, including NGC 628 (Blanc et al. 2013). We use the same local observational estimates of $\Sigma_{\rm SFR}$ for NGC 628 (hereafter $\Sigma_{{\rm SFR},obs}$) as in Shi et al. (2011) which are based on a combination of {\it GALEX} far-UV measurements (Gil de Paz et al. 2007) and {\it Spitzer} 24 $\micron$ (Kennicutt et al. 2003) and which have a $3-\sigma$ lower limit of $10^{-4}$ M$_{\odot}$ yr$^{-1}$ kpc$^{-2}$. 

For each resolution element in NGC 628 with measurements in the VENGA survey, we estimate the value of $\lambda_{\rm SF}$ by solving Eq.~\ref{eq12}. We then use Eq.~\ref{eq19} and Eq.~\ref{eq20} to evaluate the theoretical values of the SFR (${\rm SFR}_{th}$) and the surface density of the star formation rate, $\Sigma_{{\rm SFR},th}$. The number of resolution elements in NGC 628 that simultaneously have $\sigma_{*}$ measurements in VENGA as well as measured values of $\Sigma_{{\rm SFR},obs}$ is 91. Fig.~\ref{fig1} displays the distribution function of $\lambda_{\rm SF}$ for these pointings, obtained for $a=b=1/3$. The values of $a=1/3$ and $b=1/3$ are consistent with average values of these quantities derived using cold \ion{H}{i} intensity fluctuations (Lazarian \& Pogosyan 2000; Elmegreen et al. 2001; Begum et al. 2006; Dutta et al. 2009). The distribution in Fig.~\ref{fig1} peaks at $\approx 850$ pc and is positively skewed towards larger values, and we argue in App.~\ref{appb} that this result is not dependent on the spatial resolution of the observations. While there are no accurate estimates of the vertical scales heights of gas and stars in NGC 628\footnote{Kregel et al. (2002) argued that there is a constant ratio of the radial to vertical length scales in galactic disks of $l_{*}/h_{*} \approx,7.3 \pm 2.2$. With the measured value of $l_{*} \approx 2.3$ kpc in NGC 628 (Leroy et al. 2008), this yields a value of $h_{*} \approx 315$ pc, under the assumption that $h_{*}$ is independent of galactic radius. From an analysis of the \ion{H}{i} line power spectrum, Dutta et al. (2008) argued for an upper limit on the \ion{H}{i} gas vertical scales height of 800 pc.}, the values of $\lambda_{\rm SF}$ are large enough such that the condition $\lambda_{\rm SF} \gtrsim 2 \pi h_{g}$ and $\lambda_{\rm SF} \gtrsim 2 \pi h_{*}$ seems to be reasonably fulfilled for almost all resolution elements. The values of ${\rm SFR}_{th}$ and  $\Sigma_{{\rm SFR},th}$ are then derived following the formalism given in \S.~\ref{connectionsfr} with an assigned value of $\epsilon_{ff}=0.008$. A value of $0.008$ for $\epsilon_{ff}$ is consistent with the Galaxy-wide average value of $\approx 0.006$ (McKee \& Tan 2007; Murray 2011), and with the average value of $\epsilon_{ff} \approx 0.01$ found in numerical simulations (e.g., Semenov et al. 2016, see Fig.~\ref{fig2} in their paper). This value is a factor $\approx 10$ smaller than the average value measured on the scale of giant molecular clouds (GMCs) in the Galaxy (Murray 2011). This is expected since the gas is denser and more gravitationally bound in GMCs than the spatially averaged gas densities on scales of 750 pc (as are the observations of NGC 628) or on entire galactic scales. 

 \begin{figure}
\begin{center}
\epsfig{figure=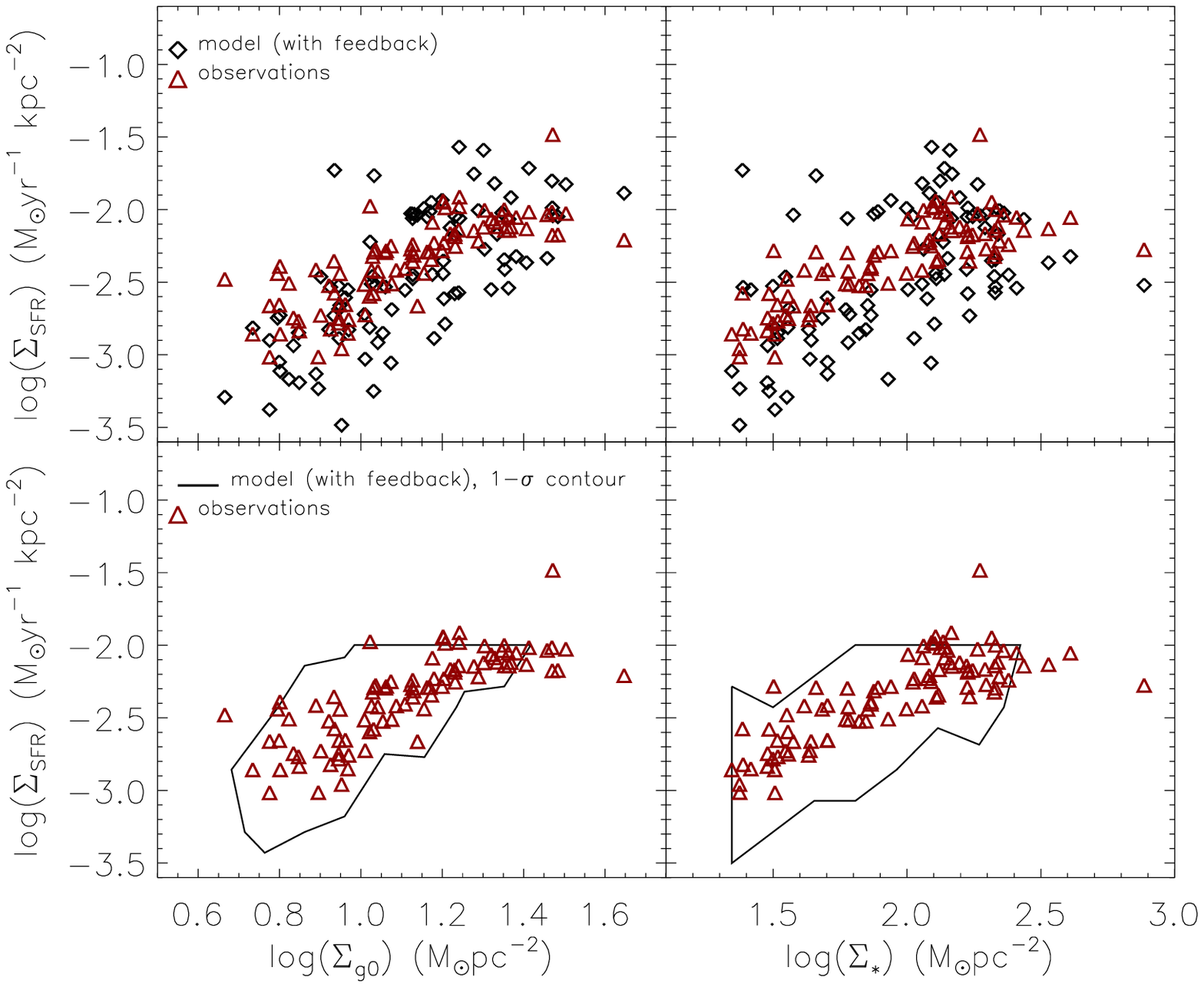,width=0.80\columnwidth}
\end{center}
\vspace{1cm}
\caption{Similar to Fig.~\ref{fig3}, but in this case the efficiency of star formation per unit free-fall time is taken to depend on the gas surface density following $\epsilon_{ff}\propto \Sigma_{g}^{-0.34}$ and normalised to be $0.8\%$ at $\Sigma_{g}=1$ M$_{\odot}$ pc$^{-2}$. Here $\Sigma_{g0}$ is the total surface density of the gas measured on a spatial scale which is equal to the spatial resolution of the observations (i.e., 750 pc).}
\label{fig5}
\end{figure}

Fig.~\ref{fig2} displays the distribution function of the ratio of the theoretical to observational star formation rates $ \left({\rm SFR}_{th}/{\rm SFR}_{obs}\right)$. The dispersion in this distribution is $\approx 0.3$ dex. Fig.~\ref{fig3} displays the scatter plots in the $\Sigma_{g}-\Sigma_{\rm SFR}$ space (left column) and in the $\Sigma_{*}-\Sigma_{\rm SFR}$ space (right column). The observations are shown with the red open triangles, and the theoretical estimates are shown with the black open diamonds (top) and as a closed contours containing $68\%$ of the theoretical points (bottom). A noticeable aspect of Fig.~\ref{fig3} is that in the low surface density regime ($\Sigma_{g} \lesssim 10-15$ M$_{\odot}$ pc$^{-2}$), the model matches perfectly the data, both in terms of the dependence of $\Sigma_{\rm SFR}$ on $\Sigma_{g}$ and $\Sigma_{*}$ and in terms of the level of dispersion at any given value of $\Sigma_{g}$ and $\Sigma_{*}$. At higher surface densities ($\Sigma_{g} \gtrsim 15$ M$_{\odot}$ pc$^{-2}$), the theoretical estimates of $\Sigma_{\rm SFR}$ are larger than the observed ones by factors of $\approx 2-5$. It is important to note that our formalism does not account explicitly for the effects of feedback from massive stars, which are more important at higher surface densities where more massive clusters can form. The increased effect of feedback at high surface densities leads to a more rapid expulsion of the gas from the clusters and to a reduction of the star formation efficiency per unit time (Dib 2011a). Dib (2011a), showed that in the surface density regime relevant for this work ($1$ M$_{\odot}$ pc$^{-2} \lesssim \Sigma_{g} \lesssim 50$ M$_{\odot}$ pc$^{-2}$), the value of the star formation efficiency per free-fall time ($\epsilon_{ff}$) decreases by a factor of $\approx 4$ going from low to higher gas surface densities, and this is valid for any given value of the gas phase metallicity. Using a scaling of $\epsilon_{ff}$ as a function of  $\Sigma_{g}$ ($\epsilon_{ff} \propto \Sigma_{g}^{-0.34}$) (Dib 2011a), and fixing the value of $\epsilon_{ff}=0.008$ at $\Sigma_{g}=1$ M$_{\odot}$ yr$^{-1}$, we make a new estimate of $\Sigma_{{\rm SFR},th}$. The distribution function of the ratio ${\rm SFR}_{th}/{\rm SFR}_{obs}$ in the presence of the effects of feedback is displayed in Fig.~\ref{fig4}. While the distribution in Fig.~\ref{fig4} does not peak at unity (because of the arbitrary choice of fixing $\epsilon_{ff}=0.008$ at $\Sigma_{g}=1$ M$_{\odot}$ yr$^{-1}$), the inclusion of a correction due to feedback removes the positive skewness of the distribution (i.e., at high surface densities) and leads to a quasi symmetric dispersion around each side of the observations. Fig.~\ref{fig5} displays the corresponding scatter plots for $\Sigma_{\rm SFR}$ versus $\Sigma_{g}$ and $\Sigma_{*}$ (left and right panels, respectively). The figure shows that the inclusion of feedback in the treatment of GI in a star+gas galactic disk is necessary in order to better match the observed dependence of $\Sigma_{\rm SFR}$ on both $\Sigma_{g}$ and $\Sigma_{*}$.
  
\section{CONCLUSIONS}\label{conclusions}

In this work, we explore the dependence of the surface density of star formation in galactic disks on the gas and stellar surface densities and velocity dispersions. We treat both gas and stars as an isothermal fluid and use the linear stability analysis of the gravitationally coupled hydrodynamical equations in order to derive the wavelength of the most unstable mode of the gravitational instability (GI) ($\lambda_{\rm SF}$). We find that the latter quantity is a function of the stellar surface density, the gas surface density, the velocity dispersion of stars, and the scaling laws of turbulence in the gas phase. When applying our model to the face-on, spiral galaxy NGC 628, for which all the required observational data are available, we find that the distribution of $\lambda_{\rm SF}$ for the ensemble of resolution elements for which the required stellar+gas data is available peaks at $\approx 850$ pc and is skewed towards higher values (with a tail of the distribution up to $\approx 2.5$ kpc; see Fig.~\ref{fig1}). Gravitational instabilities on such large scales are likely to determine the rate of giant molecular cloud (GMC) formation. In turn, stars form in GMCs with a distribution of the star formation efficiencies that depend on the distribution of GMC masses, and on the distributions of their internal physical and dynamical properties coupled to a regulation provided by stellar feedback (e.g., Padoan \& Nordlund 2011; Dib et al. 2013). It is therefore reasonable to assume that reservoirs of gas that become gravitationally unstable on large scales are correlated with the star formation rate (SFR) on these scales. For a given set of physical conditions in each resolution elements of NGC 628, we derive the theoretical value of the SFR under the assumption that the fastest growing mode of the gas+star GI is directly linked to the SFR. The theoretical surface density of the star formation rate ($\Sigma_{\rm SFR,th}$) is obtained by dividing the SFR by the physical surface area of the surface element in the observations. The only free parameters of the models are the exponents of the turbulence scaling laws of the gas (i.e., $a$, and $b$ which are the exponents of the gas surface density- and velocity dispersion size relations, see Eq.~\ref{eq10} and Eq.~\ref{eq11}), and the star formation efficiency per unit free-fall time, $\epsilon_{ff}$. The values of $a$ and $b$ and $\epsilon_{ff}$ are fixed at $a=b=1/3$ and $\epsilon_{ff}=0.8\%$, respectively. These values of $a$ and $b$ are appropriate for the description of the structure and velocity dispersion of the cold neutral hydrogen in the disk galaxies. A fixed value of $\epsilon_{ff}$ serves only as a normalisation, and does not affect neither the shapes of the $\Sigma_{g}-\Sigma_{\rm SFR}$ and $\Sigma_{*}-\Sigma_{\rm SFR}$ relations, nor the amount of scatter at any fixed value of $\Sigma_{g}$ or $\Sigma_{*}$. 

We find an encouraging match between the theoretical estimates of the surface density of star formation $\Sigma_{{\rm SFR},th}$ from our model and the observational values for NGC 628 ($\Sigma_{{\rm SFR},obs}$), both in terms of the shapes of the $\Sigma_{g}-\Sigma_{\rm SFR}$ and $\Sigma_{*}-\Sigma_{\rm SFR}$ scatter relations and in terms of the dispersion of the data points at fixed values of $\Sigma_{g}$ or $\Sigma_{*}$. The model-observations matching is further improved if the value of $\epsilon_{ff}$ is taken to decrease with increasing gas surface density as earlier suggested by Dib (2011a,b). The origin of the dependence of $\epsilon_{ff}$ on $\Sigma_{g}$ is attributed to the effects of feedback in the pre-supernova phase in stellar clusters. More massive clusters are more likely to form at higher surface densities. Gas expulsion from more massive clusters occurs on shorter timescales than in lower mass clusters (Dib et al. 2013), and the rapid expulsion of gas results in a faster quenching of star formation and to a reduction of the star formation efficiency per unit time. Our model opens a new path towards a better understanding of the dependence of the star formation rate in galaxies on the local stellar and gas properties. Higher spatial and spectral resolution observations will allow us to further constrain the model and will also help reduce the number of free parameters by directly measuring the scaling laws of turbulence.                                                                                                                                                
 
\section{Acknowledgments}

We thank the referee Alessandro Romeo for a careful reading of the paper which helped clarify some aspects of the text, Sophia Lianou and Matthew Orr for useful comments on a draft version of this paper, and Shi Yong for sharing some of the observational data. S. D. acknowledges the support provided by a Marie Curie Intra European Fellowship under the European Community's Seventh Framework Program FP7/2007-2013 grant agreement no 627008,  during the early phase of this work. S. H. acknowledges financial support from DFG program HO 5475/2-1. G. B. is supported by CONICYT/FONDECYT programa de iniciaci\'{o}n Folio 11150220. This research has made use of NASA's Astrophysics Data System Bibliographic Services.  

{}

\appendix 

\section{GOVERNING EQUATION FOR $k_{\rm SF}$ }\label{appa}

The derivation of the wavenumber of the fastest growing mode of the instability, $k_{\rm SF}$, is achieved using Eq.~\ref{eq12}, where $\omega_{-}^{2}$ is given by Eq.~\ref{eq11}. There exist an analytical expression for the general equation of $k_{\rm SF}$ which is given by: 

\begin{align}
& \sigma_{*}^{2} k_{\rm SF}+\sigma_{g0}^{2} (1-b) \left(\frac{1}{k_{0}} \right)^{-2b} k_{\rm SF}^{1-2b}- \pi G \Sigma_{*} - \pi G \Sigma_{g0} \left( \frac{k_{\rm SF}}{k_{0}}\right)^{-a}\nonumber\\
& -  {\biggr[} \left(\sigma_{*}^{2} -\sigma_{g0}^{2} \left(\frac {k_{\rm SF}}{k_{0}}\right)^{-2b} \right)^{2} k_{\rm SF}^{2} + 4 \pi G \left(\Sigma_{g0} \left(\frac{k_{\rm SF}}{k_{0}}\right)^{-a}-\Sigma_{*} \right) \left(\sigma_{*}^{2}-\sigma_{g0}^{2}\left(\frac{k_{\rm SF}}{k_{0}} \right)^{-2b}\right) k_{\rm SF}\nonumber\\ 
& + 4\pi^{2} G^{2} \left(\Sigma_{g0} \left(\frac{k_{\rm SF}}{k_{0}}\right)^{-a} +\Sigma_{*}\right)^{2}{\biggr]^{1/2}} \nonumber\\
& -  k_{\rm SF} \frac{ {\biggr[}2 k_{\rm SF} \left(\sigma_{*}^{2}-\sigma_{g0}^{2} \left( \frac{k_{\rm SF}}{k_{0}}\right)^{-2b} \right)^{2}+4 b \sigma_{g0}^{2} \left(\frac{k_{\rm SF}}{k_{0}}\right)^{-2b} k_{\rm SF} \left(\sigma_{*}^{2}-\sigma_{g0}^{2} \left(\frac{k_{\rm SF}}{k_{0}}\right)^{-2b}\right){\biggr]}} {D}\nonumber\\
& + k_{\rm SF} \frac{ {\biggr[} 4\pi G \left(\Sigma_{g0}\left( \frac{k_{\rm SF}}{k_{0}}\right)^{-a}-\Sigma_{*}\right) \left(\sigma_{*}^{2}-\sigma_{g0}^{2}\left( \frac{k_{\rm SF}}{k_{0}}\right)^{-2b}\right) - 4\pi a G \Sigma_{g0} \left(\frac{k_{\rm SF}}{k_{0}}\right)^{-a} \left(\sigma_{*}^{2}-\sigma_{g0}^{2} \left(\frac{k_{\rm SF}}{k_{0}}\right)^{-2b}\right) {\biggr]}} {D}\nonumber\\ 
& + k_{\rm SF} \frac{ {\biggr[} 8 a \pi^{2} G^{2} \Sigma_{g0} \left(\frac{k_{\rm SF}}{k_{0}}\right)^{-a} k_{\rm SF}^{-1} \left(\Sigma_{g0} \left(\frac{k_{\rm SF}}{k_{0}}\right)^{-a}+\Sigma_{*}\right) + 8 \pi b G \sigma_{g0}^{2} \left(\frac{k_{\rm SF}}{k_{0}}\right)^{-2b} \left(\Sigma_{g0} \left(\frac{k_{\rm SF}}{k_{0}}\right)^{-a}-\Sigma_{*} \right) {\biggr]}}  {D}\nonumber\\
& =  0 
\label{eqa1}
\end{align} 

with

\begin{align}
D & = 4 {\biggr[}k_{\rm SF}^{2} \left(\sigma_{*}^{2}-\sigma_{g0}^{2} \left(\frac{k_{\rm SF}}{k_{0}}\right)^{-2b} \right)^{2}+4\pi G k_{\rm SF} \left(\Sigma_{g0} \left(\frac{k_{\rm SF}}{k_{0}}\right)^{-a} - \Sigma_{*}\right) \left(\sigma_{*}^{2}-\sigma_{g0}^{2} \left(\frac{k_{\rm SF}}{k_{0}}\right)^{-2b}\right)+ \nonumber\\
& 4 \pi^{2} G^{2} \left(\Sigma_{g0} \left(\frac{k_{\rm SF}}{k_{0}}\right)^{-a}-\Sigma_{*}\right)^{2} {\biggr]^{1/2}}.
\label{eqa2}
\end{align}

Given the values of $k_{0}=2\pi/\lambda_{0}$, where $\lambda_{0}$ is the physical size of the resolution element in the observations. For each resolution element of the NGC 628 galaxy, we solve Eq.~\ref{eqa1} numerically using a globally-convergent Broyden's method (Press et al. 1992)  

\section{DO THE RESULTS DEPEND ON THE  SPATIAL RESOLUTION OF THE OBSERVATIONS ?}\label{appb}

\begin{figure}
\begin{center}
\epsfig{figure=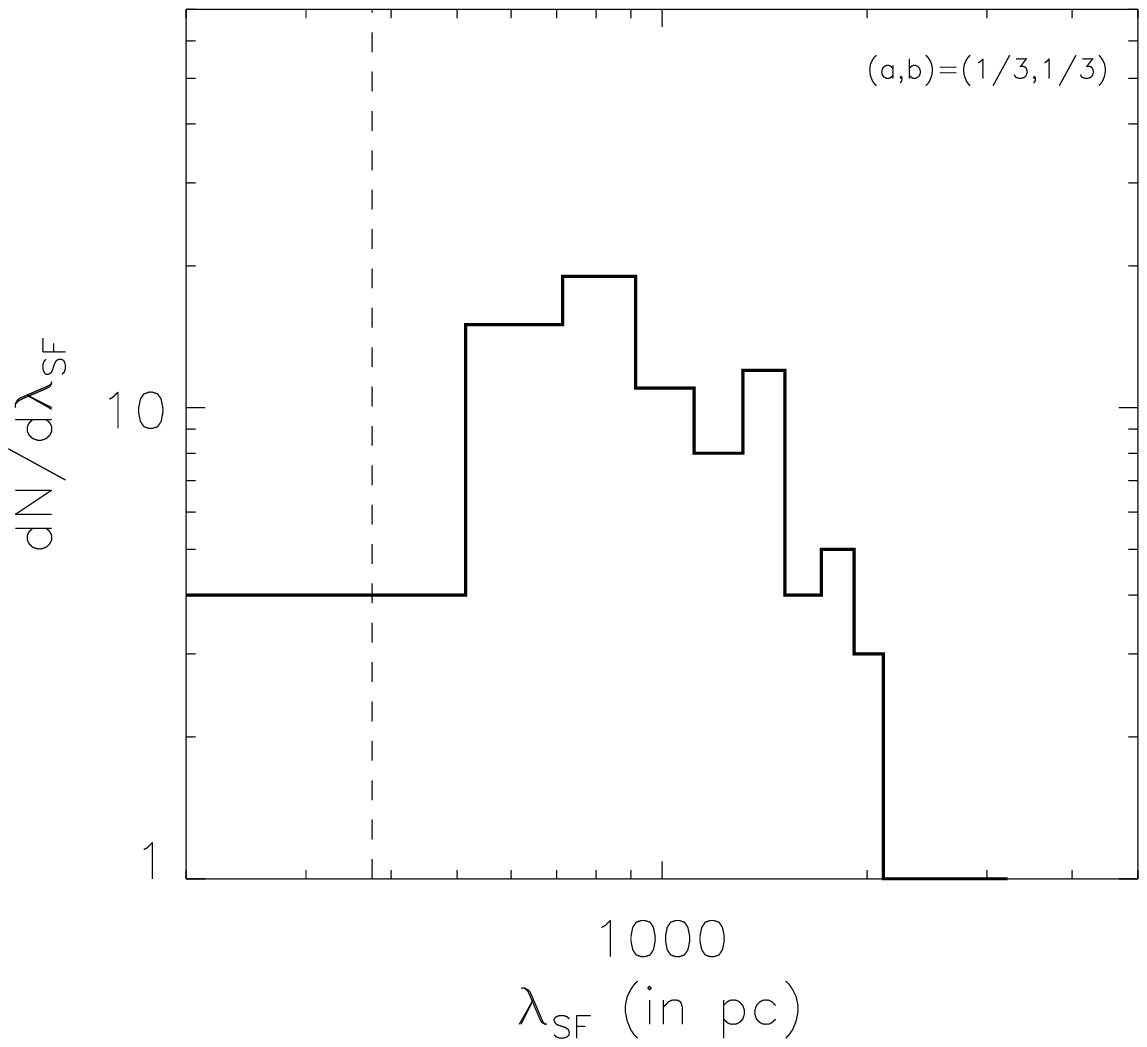,width=\columnwidth}
\end{center}
\caption{Same as Fig.~\ref{fig1} but for an adjusted spatial resolution in the observation of $375$ pc (shown as the dashed line). The values of the parameters are kept at $a=1/3$ and $b=1/3$.}
\label{fig1app}
\end{figure}

\begin{figure}
\begin{center}
\epsfig{figure=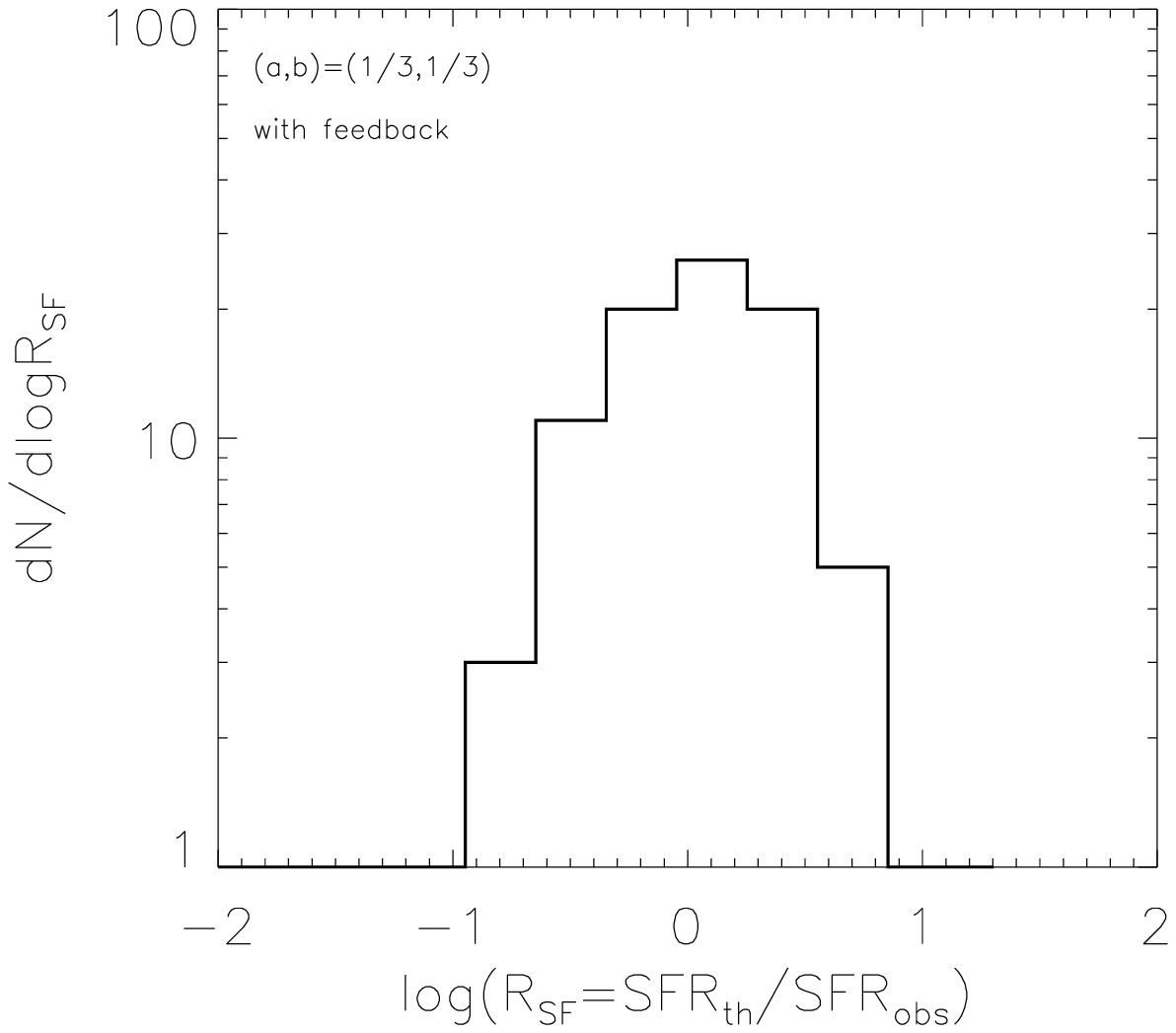,width=0.80\columnwidth}
\end{center}
\vspace{1cm}
\caption{Similar to Fig.~\ref{fig4}, but in this case we assume a spatial resolution of $375$ pc (marked in the figure by the dashed line). The stellar surface density and velocity dispersion are kept to their similar value for the resolution of $750$ pc, and the surface density and velocity dispersion of the gas are adapted using Eq.~\ref{eq10} and Eq.~\ref{eq11}, respectively.}
\label{fig2app}
\end{figure}

The question may arise whether the solutions obtained for $\lambda_{\rm SF}$ using Eq.~\ref{eq12} (i.e., Eq.~\ref{eqa1} in its detailed form) depend on the spatial resolution of the observations (here $\lambda_{0}=750$ pc). It should be noted that the surface density and velocity dispersion of the gas have a scale dependance on the dimensionless number $k/k_{0}$ (and not merely on $k_{0}$). Nonetheless, we test this by performing the following simple test. We assume that the observations have been performed on a spatial resolution of $375$ pc (thus $k_{0}$ is now replaced by $2 k_{0}$, where $k_{0}$ is the wavenumber associated with the original spatial resolution of 750 pc). We do not possess observations that have been obtained self-consistently at a spatial resolution that is half of the spatial resolution of the observations at hand. However, we adapt the current observations to present those that could be obtained with an improved spatial resolution by a factor $2$. In the absence of a better guess, the stellar surface density and velocity dispersion for the resolution $\lambda_{0}/2$ are kept the same as on the scale $\lambda_{0}$. The velocity dispersion and surface density of the gas in Eqs.~\ref{eq10} and \ref{eq11} have to be multiplied by the factors $2^{-\beta}$ and $2^{-\alpha}$, respectively. For $\alpha=\beta=1/3$, the gas velocity dispersion and surface density are both reduced by a factor $2^{-1/3}$. These assumptions generate only approximate conditions for the stellar and gas components in each constructed half-resolution element as one expects that there would be local fluctuations of the stellar velocity and surface density on smaller scales. 

\begin{figure}
\begin{center}
\epsfig{figure=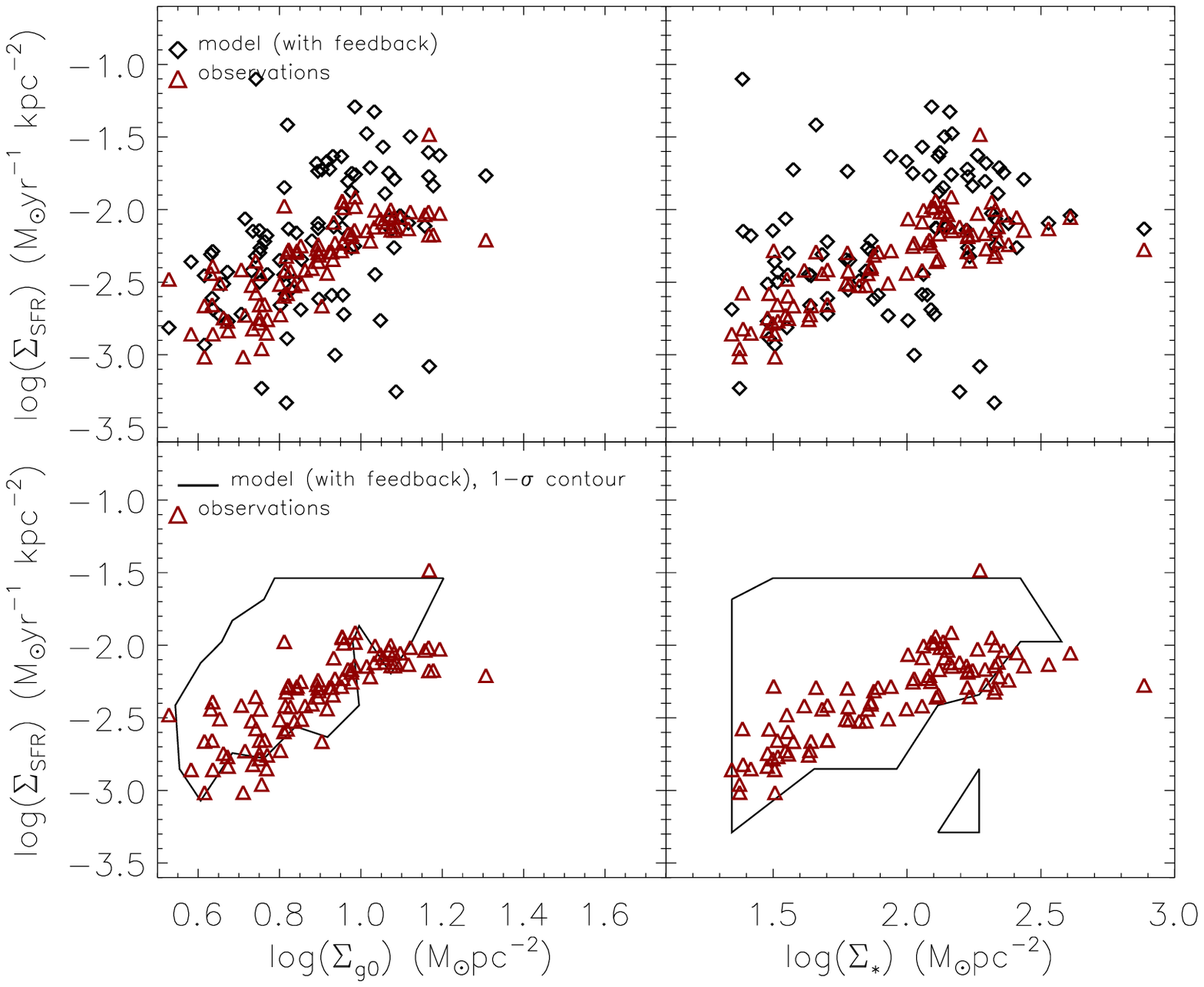,width=0.80\columnwidth}
\end{center}
\vspace{1cm}
\caption{Similar to Fig.~\ref{fig5}, but in this case we assume a spatial resolution of $375$ pc. The stellar surface density and velocity dispersion are kept to their similar value for the resolution of $750$ pc, and the surface density and velocity dispersion of the gas are adapted using Eq.~\ref{eq10} and Eq.~\ref{eq11}, respectively. Here $\Sigma_{g0}$ is the total surface density of the gas measured on a spatial scale which is equal to the adjusted spatial resolution of the observations (here 375 pc).}
\label{fig3app}
\end{figure}

Fig.~\ref{fig1app} displays the distribution of the wavelengths of the most unstable mode ($\lambda_{\rm SF}$) with the new adopted spatial resolution. As expected, the choice of a different spatial resolution (here a higher resolution) does not affect the results and the distribution of $\lambda_{\rm SF}$ still peaks at $\approx 850-900$ pc. For this same adopted spatial resolution, Fig.~\ref{fig2app} displays the ratio of the theoretical to observed star formation rates while Fig.~\ref{fig3app} displays the surface density of the star formation rate as a function of the surface density of the gas (left panels) and of the stars (right panels) (for the model as a scatter plot in the top panels and as a closed $1-\sigma$ contour in the bottom panels). In this case, the efficiency of star formation per free-fall time $\epsilon_{ff}$ has been taken to include a correction for feedback (i.e., as in Fig.~\ref{fig4} and Fig.~\ref{fig5}). The existence of more outliers which result in a larger scatter is probably due to the approximations made in constructing the physical quantities (especially $\Sigma_{*}$ and $\sigma_{*}$) for the higher spatial resolution case.
 
\label{lastpage}

\end{document}